\documentclass[twocolumn,showpacs,footnoteinbib]{revtex4}
\usepackage{amsmath,amssymb,epic,eepic}
\usepackage[dvips]{graphicx}
\usepackage{epsfig}
\usepackage{color}
\usepackage{graphicx}

\begin{document}

\title{Plasmon scattering approach to energy exchange and high frequency noise in $\nu=2$ quantum Hall edge channels}

%\today

\author{P.~Degiovanni$^{1}$}
\author{Ch. Grenier$^{1}$}
\author{G.~F\`eve$^{2}$}
\author{C.~Altimiras$^{3}$}
\author{H.~le~Sueur$^{3}$}
\author{F.~Pierre$^{3}$}

\affiliation{(1) Universit\'e de Lyon, F\'ed\'eration de Physique Andr\'e Marie Amp\`ere,
CNRS - Laboratoire de Physique de l'Ecole Normale Sup\'erieure de Lyon,
46 All\'ee d'Italie, 69364 Lyon Cedex 07,
France}

\affiliation{(2) Laboratoire Pierre Aigrain, D{\'e}partement de Physique
de l'Ecole Normale Sup\'erieure, 24 rue Lhomond, 75231 Paris Cedex
05, France}

\affiliation{(3) CNRS, Laboratoire de Photonique et de Nanostructures (LPN) - Phynano team, Route de Nozay, 91460 Marcoussis, France}

\begin{abstract}
Inter-edge channel interactions in the quantum Hall regime at filling factor $\nu= 2$ are analyzed within a plasmon scattering formalism. We derive analytical expressions for energy redistribution amongst edge channels and for high frequency noise, which are shown to fully characterize the low energy plasmon scattering. In the strong interaction limit, the predictions for energy redistribution are compared with recent experimental data and found to reproduce most of the observed features. Quantitative agreement can be achieved by assuming $25\%$ of the injected energy is lost towards other degrees of freedom, possibly the additional gapless excitations predicted for smooth edge potentials.
\end{abstract}

\pacs{73.23.-b,73.43.Cd,73.50.Td,73.43.Lp}

\maketitle

Electronic transport along the chiral edges of a two dimensional electron gas (2DEG) in the 
Quantum Hall (QH) regime 
can now be studied using a single electron source ~\cite{Feve:2007-1}, thus opening the way to
fundamental electron optics experiments such as single electron
Mach-Zenhder interferometry (MZI)~\cite{Ji:2003-1}
or Hong-Ou-Mandel experiments~\cite{olkhovskaya:2008-1}.
However, contrary to photons, electrons strongly interact through the Coulomb interaction. This leads
to relaxation and decoherence phenomena and thereby questions the whole 
concept of electron quantum optics. The $\nu=2$ filling factor is particularly appropriate to address
this issue since the electromagnetic environment of one edge channel (EC) mainly consists of the other EC.
In this respect, it is an ideal test bed to investigate interaction effects in the QH regime.

Coulomb interactions lead to plasmon scattering which in turn
basically determines the linear transport properties of one dimensional systems: 
finite frequency admittances~\cite{Safi:1999-1} and thermal conductivity~\cite{Kane:1996-1,Fazio:1998-1}.
Recently it was pointed out as a key ingredient for understanding the interference contrast of Mach-Zehnder interferometers at
$\nu=2$~\cite{Levkivskyi:2008-1} and single electron
relaxation along ECs~\cite{Degio:2009-1}.
However, despite its fundamental role in the QH regime low energy physics,
plasmon scattering has only been indirectly probed
through electron quantum interferences (MZI)~\cite{Neder:2006-1,Roulleau:2007-2,Roulleau:2008-1,Roulleau:2008-2}. 
Recent progresses in the measurements of high frequency admittance~\cite{Gabelli:2006-1,Gabelli:2007-1}, noise~\cite{Zakka-bajjani:2007-1} and
electron distribution function~\cite{Altimiras:2009-1} in these systems
open new complementary ways to probe the dynamics of QH ECs.

\medskip

\begin{figure}
\includegraphics[width=8cm]{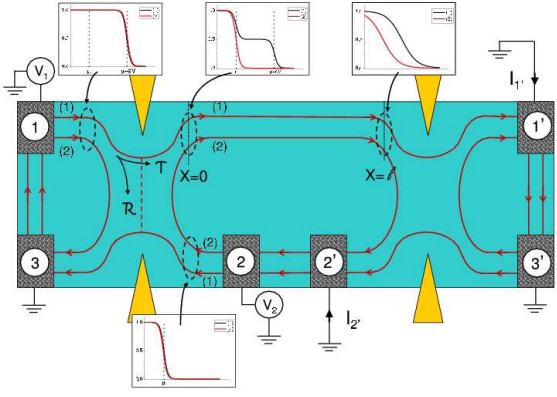}
\caption{\label{fig:schema} (Color online)
Relaxation of a non equilibrium distribution created in EC $(1)$ by a QPC of transmission $\mathcal{T}$, biased at $V$
(here $\mathcal{T}=0.5$). 
Edge channels $(1)$ and $(2)$ interact between $x=0$ and $x=L$. The electron distribution
function in channel $(1)$ evolves from a non equilibrium smeared double step to a thermal-like distribution function.
%at temperature $T_{\mathrm{eff}}(L,V,T_{\mathrm{el}},\mathcal{T})$. 
The right QPC is set to perfectly reflect (transmit) the internal (external) EC. For 
admittance measurements, the left QPC is operated as the right one. Voltage
drives $V_{1,2}(t)$ are applied and currents $I_{1',2'}(t)$  are measured at contacts $1'$ and $2'$.}
\end{figure}

In this letter, we discuss how energy relaxation and high frequency noise provide direct information on plasmon scattering
in a $\nu=2$ EC system. We anticipate it will permit us to reach a much
deeper understanding of the low energy physics of the QH regime.
The model considered here neglects possible additional excitation
modes associated with the internal structure of ECs~\cite{Chamon:1994-1,Aleiner:1994-1,Han:1997-2}. We derive
from general considerations
a universal expression for plasmon scattering that is valid at low energies. It gives the finite frequency admittances in 
the 6-terminal geometry depicted in Fig.~\ref{fig:schema} and energy exchanges
between the two ECs. The latter are compared with experimental data recently obtained from electron
relaxation in one of the edges of the $\nu=2$ system \cite{lesueur2009relaxqhr} thus providing a
quantitative test of plasmon scattering models. This comparison suggests
that our model captures most of the physics of this system. We argue that
the discrepancy between raw data and predictions might be due to energy leak in the 
predicted internal modes~\cite{Aleiner:1994-1,Han:1997-2}. 
Noise measurements in the GHz range provide
more information on plasmon scattering and regarding the limits of our model as well as  
of another approach~\cite{Lunde:2009-1} based on iterating 
collision corrections to the free electron model~\cite{Buttiker:1988-1}.

\medskip

Within the bosonization formalism, the $\nu=2$ QH ECs are
described using a two component bosonic field $\Phi=(\phi_{\alpha})$ encoding
plasmon modes of the outer ($\alpha=1$) and inner ($\alpha=2$) channels. It is related to the electronic densities by
$:(\psi_{\alpha}^\dagger\psi_{\alpha})(x):=(\partial_{x}\phi_{\alpha})(x)/\sqrt{\pi}$.
The effect of interactions between $x=0$ and
$x=L$ is conveniently described in terms of a scattering matrix $\mathbf{S}(\omega,L)$ relating
incoming ($x=0$) and outgoing ($x=L$) plasmon modes with frequency 
$\omega/2\pi$~\footnote{Plasmon scattering is elastic
because of the quadratic form of Coulomb interactions in the bosonic field $\Phi$.}.
Denoting by $\Phi(\omega,x)$ the
Fourier transform of the two component field $\Phi$ with respect to time, $\mathbf{S}(\omega,L)$ is defined as
\begin{equation}
\label{eq:coupled:S}
\Phi\left(\omega,L\right)=\mathbf{S}(\omega,L)\ldotp \Phi\left(\omega,0\right)\,.
\end{equation}
Let us now derive a universal form for $\mathbf{S}(\omega,L)$ valid at low energies.

We first consider the situation where the two QPCs shown in  Fig.~\ref{fig:schema} are set to 
fully transmit the outer EC and fully reflect the inner one. Then, 
the plasmon scattering matrix element $\mathbf{S}_{\alpha,\beta}(\omega,L)$
is related to the sample admittance~\cite{Christen:1996-1,Gabelli:2007-1} 
between contacts  $i=\alpha'$ and $j=\beta$ as depicted on Fig.~\ref{fig:schema}:
\begin{equation}
\label{eq:GfromS}
G_{i,j}(\omega,L,B_{\perp})=-\frac{e^2}{h}\mathbf{S}_{\alpha,\beta}(\omega,L)\,.
\end{equation}
Here $B_\perp$ is the applied perpendicular magnetic field.
Now, using the Onsager-B\"uttiker relations~\cite{Buttiker:1986-1} 
$G_{i,j}(\omega,L,B_{\perp})=G_{j,i}(\omega,L,-B_{\perp})$ and the mapping $(\alpha,\beta') \mapsto (\alpha',\beta)$ when 
$B_{\perp}\mapsto -B_{\perp}$ that follows from the reversed propagation direction, we 
obtain that the plasmon scattering matrix is symmetric. Energy conservation implies that $\mathbf{S}(\omega,L)$
is a unitary matrix. Consequently, $\mathbf{S}(\omega,L)$ is of the form:
\begin{equation}
\label{eq:coupled:S-general}
\mathbf{S}(\omega,L)=e^{i\theta(\omega,L)}\times e^{-i(a_{z}(\omega,L)\sigma^z+a_{x}(\omega,L)\sigma^x)}
\end{equation}
where $\theta(\omega,L)$ and $a_{x,z}(\omega,L)$ are real numbers independent of $B_{\perp}$ and $\sigma^{x,z}$ 
denote the Pauli matrices.
At fixed frequency, $\mathbf{S}(\omega,L)$ goes
to the identity for $L\rightarrow 0$. For propagation distances much larger than
the interaction range, interactions can be viewed as local and therefore, scattering
obeys $\mathbf{S}(\omega,L_1+L_2)=\mathbf{S}(\omega,L_1)\ldotp \mathbf{S}(\omega,L_2)$. Since
$\mathbf{S}(\omega,L)$ goes to the identity at
$\omega\rightarrow 0$, it is sufficient to expand
$\theta(\omega,L)$ and $a_{x,z}(\omega,L)$ at first order in $\omega$ and we finally get:
\begin{equation}
\label{eq:coupled:S:IR}
\mathbf{S}(\omega,L)=e^{i\omega L/v_0}\times e^{-i\frac{\omega L}{v}(\cos{(\theta)}\sigma^z+\sin{(\theta)}\sigma^x)}
\end{equation}
where $v_{0}$ and $v$ are velocities which, together with the angle $\theta$ completely determine
plasmon scattering. The velocities of the plasmon eigenmodes in an infinite system are  $v_{\pm}^{-1}=v_{0}^{-1}\pm v^{-1}$.
The case of uncoupled channels corresponds to $\theta\equiv 0\pmod{\pi}$ whereas any other
value corresponds to an interacting problem. 
The opposite limit $\theta\equiv\pi/2\pmod{\pi}$ corresponds, within a microscopic model 
approach~\cite{Levkivskyi:2008-1}, to either equal channel velocities
or strong inter-channel interactions.

\medskip

Besides finite frequency admittances, plasmon scattering also determines energy redistribution between the
two channels as a function of propagation distance. A plasmon of energy
$\hbar\omega$ along a distance $L$ has probability $T(\omega,L)=|\mathbf{S}_{11}(\omega,L)|^2$ to be transmitted in the
same EC and $R(\omega,L)=|\mathbf{S}_{12}(\omega,L)|^2$ to be scattered into the other channel.
The transmission probability oscillates with frequency:
$T(\omega,L)=T_{\infty}+R_{\infty}\cos{(2\omega L/v)}$ where
$T_{\infty}=1-R_{\infty}=(1+\cos^2{(\theta)})/2$. In the large $L$ limit, the energy injected in
one EC through a broadband spectrum of plasmon excitations will be redistributed according to
the coarse grained scattering probabilities $T_{\infty}$ and $R_{\infty}$. Since in the strong coupling limit,
$T_{\infty}=R_{\infty}=1/2$, this approach predicts asymptotic equipartition of
energy between the two ECs in the limit $L\rightarrow \infty$.

\medskip

To study the energy relaxation, let us consider the energy current along each EC.
It is expressed in terms of the electron distribution function $f_{\alpha}(\varepsilon,x)$ in channel $\alpha$ at position $x$ as
\begin{equation}
J_{\alpha}(x)=\int v_{\alpha}\rho_{\alpha}\,(f_{\alpha}(\varepsilon,x)-\Theta(\varepsilon))\,\varepsilon\,d\varepsilon
\end{equation}
where $v_{\alpha}$ and $\rho_{\alpha}$ respectively denote the electron velocity and
the density of states per unit of length and energy in channel $\alpha$ (in 1D systems $v_{\alpha}\rho_{\alpha}=1/h$) and
$\varepsilon$ denotes the energy difference
with the corresponding Fermi energy $\mu_{\alpha}$. 

Within the bosonization formalism, the energy current is expressed in terms of the plasmon modes occupation
numbers $\bar{n}_{\alpha}(\omega,x)$ in channel $\alpha$ at energy $\hbar\omega$ and position $x$:
\begin{equation}
\label{eq:energy-current}
J_{\alpha}(x)=\int_{0}^{+\infty}\hbar\omega\,\bar{n}_{\alpha}(\omega,x)\,\frac{d\omega}{2\pi}\,.
\end{equation}
In the setting depicted on Fig.~\ref{fig:schema}, the initial plasmon occupation numbers within both channels
are obtained from the finite frequency edge current noise injected by the QPC in each EC:
\begin{equation}
\label{eq:plasmon-numbers}
\bar{n}_{\alpha}(\omega,0)=\frac{2\pi}{e^2}\frac{(\delta S)_{i_{\alpha}(0)}(\omega)}{\omega}
\end{equation}
where $(\delta S)_{i_{\alpha}(0)}(\omega)=S_{i_{\alpha}(0)}(\omega)-\frac{e^2}{4\pi}\omega$ is the difference of
the finite frequency symmetric current noise and the zero point fluctuations.

\medskip

As shown in the top-middle inset of Fig.~\ref{fig:schema}, the outer channel is populated by a double step electron distribution
characterized by the bias voltage $V$, the QPC transmission $\mathcal{T}$ and the
temperature $T_{\mathrm{el}}$.  Consequently, 
the energy current injected into the outer channel is the sum of a thermal contribution given by~\cite{Kane:1996-1}:
\begin{equation}
\label{eq:thermal-current}
J^{(\mathrm{th})}_{\alpha=1}(T_{\mathrm{el}})=\frac{\pi^2}{6h}\,(k_{B}T_{\mathrm{el}})^2\,,
\end{equation}
and of an excess contribution associated through Eqs. \eqref{eq:energy-current}
and \eqref{eq:plasmon-numbers} with the edge current excess noise:
\begin{equation}
\label{eq:neq-noise}
S^{(\mathrm{exc})}_{i_{1}(0)}(\omega)=e^2\mathcal{R}\mathcal{T}\int_{-\infty}^{+\infty}
(\delta f)_{V,T_{\mathrm{el}}}(\omega')
(\delta f)_{V,T_{\mathrm{el}}}(\omega+\omega')\frac{d\omega '}{2\pi}\,,
\end{equation}
where  $\mathcal{R}=1-\mathcal{T}$ and $(\delta f)_{V,T_{\mathrm{el}}}(\omega)=f_{T_{\mathrm{el}}}(\hbar\omega+eV)
-f_{T_{\mathrm{el}}}(\hbar\omega)$ denotes a difference of Fermi functions at temperature $T_{\mathrm{el}}$.

\medskip

Since a thermal distribution at temperature $T_{\mathrm{el}}$ is injected into the inner channel,
the thermal part of the energy current in the outer channel is left unchanged by
propagation along a distance $L$. On the contrary, the excess
noise is attenuated by $T(\omega,L)$. This finally
leads to the excess contribution to the energy current in the outer channel at $x=L$ which we now
express as an excess temperature $T_{\mathrm{exc}}$ as in Eq.~\eqref{eq:thermal-current}:
\begin{equation}
\label{eq:T:excess}
\left(\frac{k_BT_{\mathrm{exc}}}{eV}\right)^2=\frac{3\mathcal{T}\mathcal{R}}{\pi^2}\,
\left(T_{\infty}+
R_{\infty}\frac{\mathrm{sinc}^2(L/L_{V})}{\mathrm{sinhc}^2(2\pi L/L_{\mathrm{th}}(T_{\mathrm{el}}))}\right)
\end{equation}
where $\mathrm{sinc}(x)=\sin{(x)}/x$, $\mathrm{sinhc}(x)=\sinh{(x)}/x$.
The length scales
$L_{V}= \hbar v/e|V|$ and $L_{\mathrm{th}}(T_{\mathrm{el}})=v\hbar/k_{B}T_{\mathrm{el}}$
respectively associated with the bias voltage and temperature
govern the $L$ dependence. This expression interpolates between the
initial $L=0$ value $(3/\pi^2)\mathcal{RT}$ and the asymptotic value
$(3/\pi^2)\mathcal{RT}T_{\infty}$.

\begin{figure}[tb]
\includegraphics[width=0.8\columnwidth,clip]{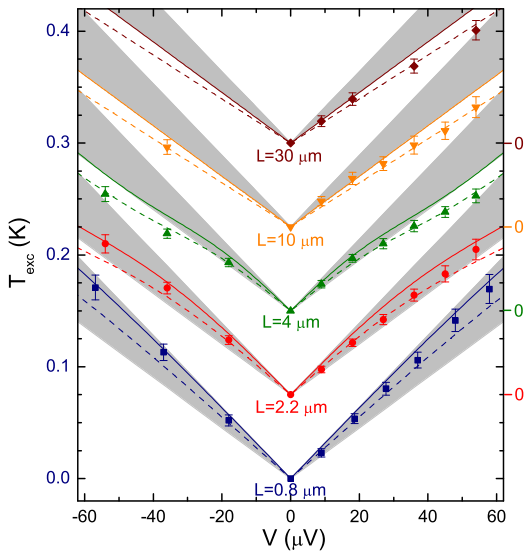}
\caption{Comparison data-prediction for $T_{\mathrm{exc}}$ plotted vs the bias voltage $V$ applied to a QPC set to $\mathcal{T}=0.5$. A 75~mK vertical shift separates consecutive $L$. Symbols are data points \cite{lesueur2009relaxqhr}. Continuous lines are predictions of Eq.~\eqref{eq:T:excess} with the known values of $T$, $V$, $L$, $\mathcal{T}$, assuming $T_\infty=1/2$ and velocity $v=10^5$~m/s. Dashed lines are the same predictions scaled down by $13\%$. Grey areas encapsulate values of $T_{\mathrm{exc}}(V,\mathcal{T}=0.5)$ accessible from Eq.~\eqref{eq:T:excess}.}
\label{Figdata1}
\end{figure}

\begin{figure}[tb]
\includegraphics[width=0.8\columnwidth,clip]{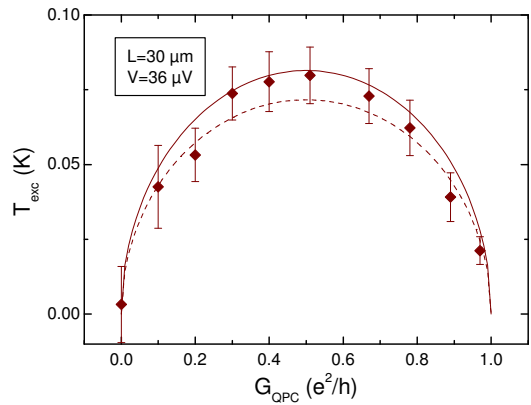}
\caption{Comparison data-prediction for $T_{\mathrm{exc}}(\mathcal{T})$. Symbols are data points vs $G_{QPC}=\mathcal{T}e^2/h$ obtained at $L=30~\mu$m and $V=36~\mu$V in a different run than \cite{lesueur2009relaxqhr}, after a thermal cycle to room temperature. The continuous line is the $L=\infty$ prediction of Eq.~\ref{eq:T:excess} with $T_\infty=1/2$. The dashed line is the same prediction scaled down by $13\%$.}
\label{Figdata2}
\end{figure}

We now confront our quantitative predictions with experimental data. In the very recent experiment \cite{lesueur2009relaxqhr} performed on a typical GaAs/Ga(Al)As semiconductor heterojunction set to Landau level filling factor 2, $T_{\mathrm{exc}}$ is extracted from measurements of the electronic energy distribution function. First, Fig.~\ref{Figdata1} shows as symbols $T_{\mathrm{exc}}$ plotted vs the QPC bias $V$ for a fixed transmission $\mathcal{T}=0.5$. In the strong interaction limit $T_\infty=1/2$, and using the velocity $v=10^5$~m/s,
eq.~\eqref{eq:T:excess} reproduces the characteristic energy relaxation length as well as the observed non-linear shape of $T_{\mathrm{exc}}(V)$, most pronounced for $L=4~\mu$m. A quantitative discrepancy remains at $L \geq 4~\mu$m, which can not be accounted for within our theoretical framework as data points are outside the grey areas.
This suggests that additional degrees of freedom are involved, such as internal modes of ECs \cite{Aleiner:1994-1}. Energy transfers to these modes
was not ruled out by experiment for the inner EC, expected wider and therefore more
prone to such phenomena than the outer EC. We find a good quantitative agreement with the data at $L \geq 4~\mu$m by applying a $13\%$ reduction to the predicted $T_\mathrm{exc}$, which corresponds to the absorption of $25\%$ of the injected energy by additional degrees of freedom (dashed lines). Second, figure~\ref{Figdata2} shows as symbols $T_{\mathrm{exc}}$ vs $\mathcal{T}$ obtained after a large energy relaxation, at $L=30~\mu$m. We find that the predicted proportionality of $T_{\mathrm{exc}}(\mathcal{T})$ with $\sqrt{\mathcal{T}(1-\mathcal{T})}$ is obeyed. Note that in this experimental run different 
from~\cite{lesueur2009relaxqhr}, the larger uncertainty would hide the $13\%$ discrepancy observed above.

\medskip

To get a better understanding of this discrepancy, it would be very useful to measure the frequency dependance of the plasmon
transmission probability $T(\omega,L)$ as well as the probability $R(\omega,L)$
for a plasmon to be scattered from one edge to the other. 
This can be achieved through high frequency noise measurement which are now available within the
GHz domain and with a typical bandwidth down to 50 MHz~\cite{Zakka-bajjani:2007-1}. 

\begin{figure}
\includegraphics[width=1\columnwidth,clip]{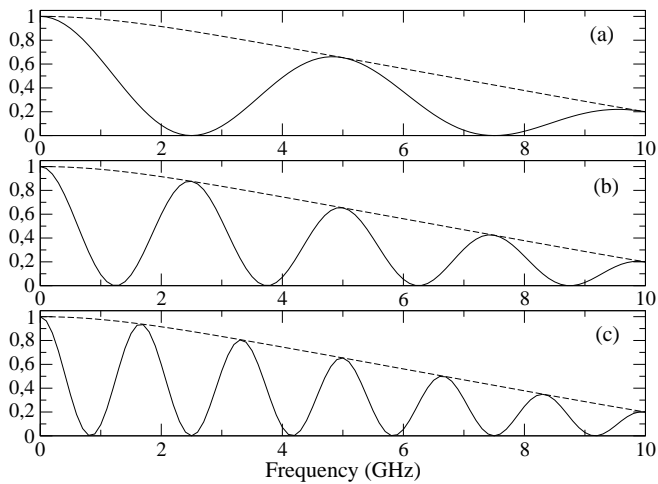}
\caption{Prediction for the frequency dependence of $S_{i_{1}(L)}^{(\mathrm{exc})}(\omega)/S_{i_{1}(L)}^{(\mathrm{exc})}(\omega=0)$ in the $0-10$~GHz range for $V=50~\mu\mathrm{V}$ and (a) $L=10$, (b) $20$ and (c) $30~\mu\mathrm{m}$. The electronic temperature is fixed to $T_{\mathrm{el}}=30~\mathrm{mK}$. We have assumed
$v=10^5~\mathrm{m}\mathrm{s}^{-1}$ and $\theta=\pi/2$. Dashed lines show the non equilibrium excess noise produced by the QPC (no relaxation).
\label{fig:noise}}
\end{figure}

The finite frequency excess noise for the currents $I_{1'}$ and $I_{2'}$ entering Ohmic contacts gives access to
the edge current excess noises $(S_{i_{\alpha}(L)}^{(\mathrm{exc})}(\omega))_{\alpha=1,2}$ since
$S_{i_{\alpha}(L)}^{(\mathrm{exc})}(\omega)=S_{I_{\alpha'}}(\omega,V)-S_{I_{\alpha'}}(\omega,V=0)$
for $\alpha=1,2$. All ohmic contacts being at temperature $T_{\mathrm{el}}$, 
after a distance $L$, the symmetric edge current excess noises read:
\begin{eqnarray}
S_{i_{1}(L)}^{(\mathrm{exc})}(\omega) & = &  T(\omega,L)\,S^{(\mathrm{exc})}_{i_{1}(0)}(\omega)\\
S_{i_{2}(L)}^{(\mathrm{exc})}(\omega) & = & R(\omega,L)\,S^{(\mathrm{exc})}_{i_{1}(0)}(\omega)\,,
\end{eqnarray}
where $S^{(\mathrm{exc})}_{i_{1}(0)}(\omega)$ is given by \eqref{eq:neq-noise}.
Using the low energy scattering matrix \eqref{eq:coupled:S:IR}, oscillations as a function of frequency are expected. Accessing 4 to 8~GHz frequencies requires $|V|\geq 50~\mu\mathrm{V}$ and
$L\geq 20~\mu\mathrm{m}$ to exhibit at least two oscillations. Fig. \ref{fig:noise} depicts
$S_{i_{1}(L)}^{(\mathrm{exc})}(\omega)/S_{i_{1}(L)}^{(\mathrm{exc})}(0)$ calculated 
for various $L$ at fixed $V$, assuming $v\simeq 10^5~\mathrm{m}\mathrm{s}^{-1}$ 
and $\theta=\pi/2$. The zero frequency excess noise
$S_{i_{1}(L)}^{(\mathrm{exc})}(0)=e^2\mathcal{R}\mathcal{T}\frac{eV}{h}\left(
\coth{(eV/2k_{B}T_{\mathrm{el}})}-2k_{B}T_{\mathrm{el}}/ eV\right)$ is left unaffected by
interactions since $\mathbf{S}(\omega=0,L)=\mathbf{1}$.

\medskip

In this Letter we have addressed the issue of accessing plasmon scattering in $\nu=2$ edge states through energy
exchange measurement and finite frequency noise. We have argued that a universal
low energy plasmon scattering matrix can be derived for systems with screened Coulomb interactions.
Comparing predictions obtained from this matrix with recent experimental data, we argue that
this approach adequatly captures most of the physics of energy exchange in this system and discuss
possible explanations~\cite{Chamon:1994-1,Aleiner:1994-1,Han:1997-2} 
to the small discrepancy between experimental data and our  model.
We then argue that high frequency noise provides a direct probe of plasmon scattering and that its measurement
would lead to a deeper understanding of the validity and limits of the plasmon model. Finally, let us stress that
sample design could be used to modulate inter channel interactions, thus leading to the realization of plasmon 
beam splitters, a building block for magnetoplasmon quantum optics.

{\bf Note:} During completion of this work we became aware of related work by A.~M.~Lunde {\it et al.} \cite{Lunde:2009-1}
based on the iteration of a collision approach for short distance equilibration by Coulomb interactions. 

\acknowledgements{We warmly thank F.~Portier for useful discussions on high frequency noise measurement.}

\end{document}